\documentclass[12pt]{article}
\usepackage{amsfonts}
\newcommand{\ba}{\begin{array}}

\newcommand{\ea}{\end{array}}

\newcommand{\be}{\begin{equation}}

\newcommand{\ee}{\end{equation}}
\newcommand{\nn}{\nonumber}

\newcommand{\bea}{\begin{eqnarray}}
\newcommand{\ena}{\end{eqnarray}}
\newcommand{\beas}{\begin{eqnarray*}}
\newcommand{\enas}{\end{eqnarray*}}
\newcommand{\mb}{\mbox}

\newcommand{\vvv}{\vec \varepsilon}

\newcommand{\vv}{\varepsilon}

\begin{document}
\pagestyle{empty}

\begin{center} 
  {\Large \textsf{
      3D Ising Model on Dual BCC Lattice:
      \\[4mm]
      the Sign-Factor}} \\
\vspace{36pt}
{\bf J.~Ambjorn \footnote{e-mail:{\sl ambjorn@alf.nbi.dk}}},
\\
Niels Bohr Institute, Blegdamsvej 17, Copenhagen, 2100, Denmark
\\
{\bf Sh.~Khachatryan \footnote{e-mail:{\sl shah@moon.yerphi.am}}},\,
{\bf A.~Sedrakyan \footnote{e-mail:{\sl sedrak@lx2.yerphi.am }}},
\\
Yerevan Physics Institute, Br.Alikhanian str.2, Yerevan 36, Armenia

\end{center}
\vfill
\begin{center}

{\bf Abstract}
\end{center}
\vspace{1cm}
The 3d Ising model on a regular cubic lattice can be expressed 
in terms of an $SU(2)$ 2d fermionic model with a $Z_2$-fluxes.
We modify the model such that it is defined on the dual to a 
body centered cubic lattice. The advantage of this lattice is that 
2d embedded surfaces have no selfintersections, thus partially
avoiding the Sign-factor problem associated with the 
2d fermionoc models related to the 3d Ising model. Rather than solving 
the full $SU(2)$ fermionic theory on this lattice we study the 
simpler model of scalar fermions and find the spectrum of excitations.
The model has no mass gap. We reformulate the model using the 
$R$-formalism and a new interesting structure appears due to the 
necessity of introducing a three-paticle matrix $R^{(3)}_{ijk}$.
It encodes the essential character of the Sign-factor. 
We analyse the integrability properties of this class of models.

\newpage
\pagestyle{plain}
\setcounter{page}{1}

\section{Introduction}
\indent

After a conjecture by A.Polyakov \cite{P} that the three-dimensional
Ising Model (3DIM) can be represented as a fermionic string theory
in a three dimensional Euclidean space a number of works was carried out
during the eighties 
\cite{PD,FSS,I,SS,CFW,AS1,AS2,AS3, PO} trying to 
substantiate the conjecture.
The main problem is the so called Sign-factor, which is
the analogy of the Kac-Ward factor \cite{KW} for the Ising Model
formulated on a 2D regular lattice. This Sign-factor
 ensures the cancellation of self-intersecting 2D surfaces mapped
onto the  target space  3D regular lattice. The cancellation is necessary
in order to have correct Boltzmann weights in the 
string representation of the 3D statistical sum as an ensemble of the
random 2D surfaces \cite{PD, FSS, CFW, AS1, AS2}. It resembles
the Pauli principle for Fermi particles, which prohibits two of them
to be in the same state (in the same space-time point in this case).
Therefore it is expected that the Sign-factor is connected
with some fermionic structure on the 2d world-sheet
of the strings.

In  \cite{AS2} a fermionic model with the Sign-factor was defined
 on the 2D regular lattice, and its 
continuum limit was investigated in \cite{AS3, AS4}. The model 
appeared to be a model of SO(3) fermions hopping in the
staggered field of $Z_2$-fluxes on a random Manhattan lattice.

A similar model of fermions hopping in a staggered $U(1)$-gauge 
field background on a regular Manhattan lattice was analysed
in \cite{AS4}.

In  \cite{CCV} a model of spins  with $Z_2$
global symmetry was considered on the so called dual Body Centered Cubic
(BCC) lattice (see Fig.1). The BCC lattice consists of two simple cubic 
sublattices arranged in such a 
way that the sites of one sublattice are positioned in the 
middle points of the cubes of the second sublattice (see Fig.1a). 

The dual to the BCC (DBCC) lattice is represented in the Fig.1b. 
As one can see,   
the two dimensional faces of the dual BCC lattice are:\\
\noindent
i) hexagons, which are dual
to the links connecting the neighbour vertices of different sublattices 
(dotted lines in 
the Fig.1b) and\\
\noindent
ii) squares, which are orthogonal to the links of the same sublattice 
(bold lines in the Fig.1b). 

This lattice is interesting in a sense
that the relevant surfaces here, i..e.\ those 
which at most  occupy the faces of the
DBCC lattice once,  have no self-intersections
at all. The analogy for contours on a 2D lattice is the honeycomb
(or its dual triangular) lattice, where curves  have no 
self-intersection points.

We follow the same reasoning as in \cite{AS2} in the case of a regular
lattice, but now on a DBCC lattice. We then obtain a fermionic 
hopping model in a staggered $SU(2)$-gauge field, defined on the 
Manhattan lattice of the type shown in Fig.\ 3.
On this lattice we do not have the magnetic fluxes which 
represented the ends of the self-intersection lines on the 
cubic lattices. However, we will see the emergence of a new 
structure of the action, which encodes the essential character 
of the Sign-factor and which is interesting from the point 
of view of integrability of the model.

In this article we define and investigate a model of fermions
hopping in a staggered $U(1)$-gauge field background on
the ML  defined in Fig.3. We find the spectrum
of the model and analyse its continuum limit. We introduce a three particle
scattering $R^{3}$-matrix, which, together with the ordinary two particle 
$R^{2}$-matrix constitutes the Monodromy matrix of the model by 
appearing alternating in a product.
We analyse the integrability property of this type of model and write down
the corresponding Yang-Baxter equations (YBE).

\section{The Model}
\indent

We consider now a simple 2d fermionic system hopping on the ML  
(Fig.3) which  was
constructed on particular  surfaces on the  DBCC lattice as 
mentioned above. This simplification
will allow us to analyse the continuum limit of the model and
investigate integrability structure.

The ML structure defines the hopping of fermions only along the arrows of the 
lattice (see Fig.\ 3) and the Hamiltonian is non hermitian:

\be
\label{H1}
 H = \sum_{<\vec n,\vec m>} t_{\vec n,\vec m}c^+
(\vec n)U_{\vec n,\vec m}c(\vec m) + \sum_{\vec n} c^+(\vec n)c(\vec n).
\ee

Here $U_{\vec n,\vec m}$ are the group elements of the external $U(1)$ 
field defined according to demand that the magnetic fluxes around the all
plaquettes with circulating arrows will be $ \phi=2\pi p/q $ 
(p and q are mutually
prime integer numbers). The hopping parameters $t_{\vec n, \vec m}$ are
chosen to be periodic, reflecting the translational symmetry of the lattice 
by vectors $2\vec i_x$, $2(1+\sqrt3)\vec i_y$,where $ \vec i_x$ 
 and $\vec i_y$ are unit 
vectors in coordinates direction.
It allows us to distinguish 10 types of particles (Fig.1):

In order to diagonalize the Hamiltonian (\ref{H1}) we pass to 
the Bloch wave basis:

\beas
c_i(\vec k)&=&\frac{1}{\sqrt{2\pi N_xN_y}}\sum_{n_x,n_y}
e^{-i\vec k(\vec n+\vec {r_i})}c_i(\vec n),\\
c_i^+(\vec k)&=&\frac{1}{\sqrt{2\pi N_xN_y}}\sum_{n_x,n_y}
e^{i\vec k(\vec n+\vec{r_i})}c_i^+(\vec n),\\
k_{x(y)}&=&\frac{2\pi\mb {\sl n}_{x(y)}}{N_{x(y)}},\quad{ }
\mb {\sl n}_{x(y)}=0,...,N_{x(y)} -1.
\enas

($ N_x,2N_y$ are the numbers of the hexagons in the corresponding directions),
and the Hamiltonian  becomes:

\be
 H(t_{i.j},\Phi)=
\sum_{\vec k}c_i^+(\vec k)\mb{\sl H}_{i,j}(\vec k,\Phi)c_i(\vec n),
\ee
where $\mb \sl H(k_x, k_y)$ is a $10\times 10$ matrix:

$$\left(
\begin{array}{cccccccccc}
\sb 1&\sb{t_{12}e^{ip}}&\sb0& \sb{t_{14}e^{iq}}& \sb0& \sb0& \sb0&\sb0& \sb0& 
\sb0\\
\sb{t_{21}e^{-ip}}&\sb 1&\sb0&\sb0&\sb0&\sb0&\sb0&\sb0&\sb0&
\sb{t_{20}e^{\frac{-iQ}{2}}}\\
\sb0&\sb{t_{32}e^{-iq}}&\sb1&\sb{t_{34}e^{-ip}}&\sb0&\sb0&\sb0&\sb0&\sb0&\sb0\\
\sb0&\sb0&\sb{t_{43}e^{-ip}}&\sb1&\sb{t_{45}e^{iQ}}&\sb0&\sb0&\sb0&\sb0&\sb0\\
\sb0&\sb0&\sb{t_{53}e^{\frac{iP}{2}}}&\sb0&\sb1&\sb{t_{56}e^{iQ}}&\sb0&\sb0&
\sb0&\sb0\\
\sb0&\sb0&\sb0&\sb0&\sb0&\sb1&\sb{t_{67}e^{-ip}}&\sb0&\sb{t_{69}e^{iq}}&\sb0\\
\sb0&\sb0&\sb0&\sb0&\sb{t_{75}e^{\frac{iP}{2}}}&\sb{t_{76}e^{-ip}}&\sb1&\sb0&
\sb0&\sb0\\
\sb0&\sb0&\sb0&\sb0&\sb0&\sb0&\sb{t_{87}e^{-iq}}&\sb1&\sb{t_{89}e^{ip}}&\sb0\\
\sb0&\sb0&\sb0&\sb0&\sb0&\sb0&\sb0&\sb{t_{98}e^{ip}}&\sb1&\sb{t_{90}
e^{\frac{-iP}{2}}}\\
\sb{t_{01}e^{\frac{-iP}{2}}}&\sb0&\sb0&\sb0&\sb0&\sb0&\sb0&\sb{t_{08}e^{\sb
\frac{-iQ}{2}}}&\sb0&\sb1

\end{array}
\right)
$$

We have defined $k_x=p$, $\sqrt3 k_y=q$, $p+q=Q$ and $p-q=P$.

 The partition function for the excitations of the states with energy $E$ 
 is the functional integral over the Grassmann variables
$\psi_i(\vec n),\bar \psi_i(\vec n)$ corresponding to the fermions.

\be
\label{Z1}
Z(E)=tre^{-\beta H}=\int \prod d\psi _i(\vec n)d
\bar \psi _i(\vec n)e^{-H(\psi,\bar \psi,\Phi) + E \bar\psi \psi}.
\ee

Since this integral is Gaussian with respect to 
$\psi _i(\vec n),\bar\psi _i(\vec n)$
fields it follows that 
\be
Z(E)=\prod_{k_x, k_y} \left[\det{\mb\sl H}(k_x, k_y)-E\right].
\ee
  
The equation $ \det{\mb\sl H}(k_x, k_y)=0 $ gives the critical line
for the low lying excitations. 

The investigation of the excitations  is conveniently
done by passing from the Lagrangian (action) formalism in the $(1+1)$ 
Euclidean space-time (formula \ref{Z1}) to the Hamiltonian and consider the 
representation of the partition function via the Transfer Matrix as follows

\be
\label{WW}
Z(0)=tr\lbrack T \rbrack^{N_y }.
\ee

By analysing the  periodic structure of the $ML$ under consideration, 
the complete Transfer Matrix can be written 
as the product of  four transfer matrices $T_i$, $i=1,2,3,4$ each of
which  mediates the  evolution
between the states $|t+i-1\rangle$ and $|t+i\rangle$, respectively, (Fig.\ 3):
\begin{equation}
\label{1T}
\mid t+i\rangle = T_i \mid t+i-1 \rangle
\end{equation}
and the complete Transfer Matrix is equal to
\be
\label{T4}
T=T_1T_2T_3T_4.
\ee

The technique of passing from 
the Lagrangian to the Hamiltonian for models on $ML$
was developed in  \cite{AS4}, based on the concept of coherent
states \cite{coh}. It is possible to introduce two types of coherent
states which are the eigenstates of the creation or annihilation 
operators, respectively. As  shown in \cite{AS4} for a $ML$,
we need to consider  two types of coherent sates alternating in a chain.
One state differs from other by particle-hole transformation.
Consequently we will need a particle and hole ordering
prescription in order to complete the definition of the Transfer Matrices.
Let us consider particle $|\psi\rangle$ and hole $|\bar\psi\rangle$ states as
follows

\bea
|\psi _{2i}\rangle&=&e^{\psi^{2i}c^+_{2i}}|0\rangle,\hspace{25mm} 
\langle\bar \psi_{2i}|=\langle0|e^{c_{2i}\bar \psi_{2i}} ,\nn\\
|\bar \psi_{2i+1}\rangle&=&(c^+_{2i+1}-\bar \psi_{2i+1})|0\rangle,\hspace{5mm}
\langle\psi _{2i+1}|=\langle0|(c_{2i+1}+\psi _{2i+1}).\nn\\
\langle\bar \psi _{2i}|\psi _{2i}\rangle&=&e^{\bar \psi _{2i}\psi 
_{2i}},\hspace{15mm}
\langle\psi _{2i+1}|\bar\psi _{2i+1}\rangle=e^{\psi _{2i+1}\bar\psi_{2i+1}}
\label{coherent}
\ena
with the properties
\bea
\label{cohb} 
 \int d\bar \psi_{2i} d\psi_{2i} |\psi _2i\rangle\langle\bar \psi _2|e^{\bar 
\psi_{2i} 
\psi_{2i} }=1,\nn\\
  \int d\bar \psi_{2i+1} d\psi_{2i+1}|\bar \psi_{2i+1}\rangle
\langle\psi _{2i+1}|e^{\bar \psi_{2i+1} \psi_{2i+1}}=1.
 \ena

 As it is clear from Fig.\ 3, the Transfer Matrices $T_2$ and 
$T_4$ are permuting the disposition of particle and 
hole type of coherent states  
at the odd and even sites, and because of this  the definition 
of the coherent state of the whole 
chain at the Euclidean times $|t+2\rangle,\quad 
|t+3\rangle$ 
must be interchanged.

Following  \cite{AS4} it is easy to see that the Transfer Matrices
  $ T_i $ are of the form :
\be
\label{Hi}
T_i= :\exp{ -H_i}:,
\ee
\bea
\label{4H}
H_1 &=& \sum_i\lbrack -t_{12}c^+_1(2i-1)c_2(2i)+ t_{34}c^+_2(2i)c_1(2i-1)-
\nn\\
&-&(t_{14}+1)c^+_1(2i-1)c_1(2i-1)+(t_{32}+1)c^+_2(2i)c_2(2i)\rbrack,\\
H_2 &=& \sum_i\lbrack -t_{43}c^+_1(2i+1)c_2(2i)+t_{45}t_{53}c^+_1(2i-1)
c_2(2i)+t_{76}c^+_2(2i+1)c_1(2i)-\nn\\ 
&-&t_{65}t_{57}c^+_2(2i-1)c_1(2i)-(1-t_{56}t_{45})c^+_1(2i)c_1(2i-1)+\nn\\
&+&(1-t_{75}t_{53})c^+_2(2i)c_2(2i-1)\rbrack,\\
 H_3 &=& \sum_i\lbrack -t_{67}c^+_1(2i)c_2(2i-1)+ t_{89}
c^+_2(2i-1)c_1(2i)+\nn\\
&-&(t_{69}+1)c^+_1(2i)c_1(2i)+(t_{87}+1)
c^+_2(2i-1)c_2(2i-1)\rbrack,\\
H_4 &=& \sum_i\lbrack -t_{98}c^+_1(2i)c_2(2i+1)+t_{90}
t_{08}c^+_1(2i)c_2(2i-1)+t_{21}c^+_2(2i)c_1(2i+1)-\nn\\
&-&t_{20}t_{01}c^+_2(2i)c_1(2i-1)+(1-t_{90}t_{0,1})
c^+_1(2i)c_1(2i-1)+\nn\\
&+&(1-t_{20}t_{08})c^+_2(2i)c_2(2i-1)\rbrack,
\ena
where under the simbol $::$ we mean ordinary normal ordering
of the fermionic operators at the even sites of the chain and
anti-normal ordering for the odd sites.

Evaluating now the product $T_1 T_2 T_3 T_4 $ for the whole 
Transfer Matrix $T$ (\ref{T4})
in the basis of coherent states  and puting it into the formula (\ref{WW}),
one can obtain the expression (\ref{Z1}) for the partition function
$Z$.

Making a Fourier transformation of the 
Hamiltonians (\ref{Hi}) and  expressing the product (\ref{T4}) of the 
Transfer matrices by use of the
generators of the $ sl_2 $ algebra in a Schwinger form
\bea
\label{1H}
  T&=&De^{-H}=D\prod_p e^{-H_p}, \label{T}\\
  H_p&=&\vvv \vec {S_p} +\mu(n_{p1}+n_{p2}),\\
  S_1&=&c^+_1c_2-c_2^+c_1,\hspace{3mm}S_2=c^+_1c_2+c_2^+c_1,\\
  S_3&=&c_1^+c_1-c_2^+c_2,\hspace{3mm}S_0=I,
\ena
one can, after some algebra, obtain the spectrum of the excitations.
In order to do that, and for simplicity, from now on we will consider two 
particular types of 
parameterization of the hopping amplitudes: 

i)
\bea
 t_{12}&=&t_{14}=t_{34}=t_{32}=t_{67}=t_{69}=t_{87}=t_{76}=t_{89}
=t_{98}=t_{21}=t_{43}=t, \nonumber\\
 t_{53}&=&t_{56}=t_{45}=t_{75}=t_{08}=t_{90}=t_{20}
=t_{01}=t^{\frac{1}{2}}
\ena
and

ii)
\be
\ba{l}
t_{i5}=t_{5j}=t_{i0}=t_{0j}=t,\\
t_{ij}=t_1,\hspace{5mm} i,j \not= 5,0,\hspace{5mm}t_1/t=x.
\ea
\ee
The first parameterization is chosen by simplicity arguments
while the second is demanded by the rotation invariance.

First we will analyse the case $i)$. The eigenvalues of the Hamiltonian
(\ref{1H}) is found to be $E=\pm \varepsilon(p)$ with the following 
dispersion relation, chemical potential $\mu$ and the parameter
$D$
\begin{eqnarray}
\label{ttt}
&&\begin{array}{ll}
\cosh
\varepsilon(p)&=2\sin^2\frac{\Phi}{2}\left(\left(1+4\cos^2\Phi 
\right)t^4+4\cos\Phi t^2-2\cos\Phi \right)+2
+1/2t^{-4}+2\cos\Phi t^{-2}\nonumber\\
&+2\cos 2p\left( 2\sin^2\frac{\Phi}{2}\left((2\cos\Phi-1)t^2-2\cos\Phi 
t^4+
1\right)-2\cos\Phi-t^{-2}  \right)\nonumber\\
&+\cos 4p\left(1-4\sin^2\frac{\Phi}{2} t^2\right),
\end{array}\\
&&\begin{array}{ll}
\mu&=0,\\
 D&=t^{4N_x}.
\end{array}
\end{eqnarray}

The condition when the massless excitations appear is defined by the equation
\begin{equation}
\label{cr}
\begin{array}{l}
2\sin^2\frac{\Phi}{2}\left((1-8\sin^2\frac{\Phi}{2}\cos\Phi)t^8+(8\cos\Phi-4) 
t^6+4\sin^2\frac{\Phi}{2}t^4\right)
+\frac{1}{2}= \\ 
=2t^4(2\cos\Phi-1)+4\sin^2\frac{\Phi}{2}t^{2}.
\end{array}
\end{equation}

The limit $t \to \infty,\quad \Phi \to 0$ on the critical line (\ref{cr}) 
becomes
\be
\sin{\frac{\Phi}{2}}t^2 =1,
\ee
and the dispersion relation  simplifies 
to 
\be
\cosh\vv=12(1-\cos 2p)+ \cos 4p.
\ee

We see that at the point p=0 there is no gap in the spectrum $\vv(0)=0$, 
and the
excitations energy near that point is a linear function of the
momentum:

\be
\vv(p)=\pm 4\sqrt{2}p.
\ee

In the case $ii)$ the model under discussion here is connected with the
Chalker-Coddington phenomenological model \cite{AS4,CC} for the 
edge excitations 
in Hall effect responsible for the plateau-plateau transitions. 

 The equation of the spectrum in this case is the following
 \be
\ba{ll}
\hspace{5mm}&\cosh \varepsilon(p)=1+(\cos{4p}-1)(t^{-2}+2(\cos\Phi-x^2)t)+\\
&2(\cos{2p}-1)\left(xt^3\cos\Phi(2x^2\cos\Phi-1-x^4)+\right.\\
&\left.-2t^2x^{-1}\cos\Phi(\cos\Phi-x^2)+x(x^2-\cos\Phi)-t^{-1}x{-1}\cos\Phi-
t^{-4}x^{-1}\right),
\ea 
\ee
and the 
equation of the critical line takes the form
\be
\ba{l}
(x^4+1-2x^2\cos\Phi)x^2\left(2t^{10}\cos\Phi-2x\cos\Phi t^9+x^2t^8/2 \right)\\
+(x^2-\cos\Phi)\left(4t^8x\cos\Phi-2t^7x^2-t^6+2t^5x\right)
+(1-x^2\cos\Phi)\left(4t^7x^2\cos\Phi-2t^6x^3\right)\\
+(\cos\Phi^2-x^2)t^6=2t^5x\cos\Phi-t^4x^2+2t^2x.
\ea
\label{cr1}
\ee

It is easy to see from this equation that for the fully packed phase:
$ t_{i,j}\rightarrow \infty$, there are two
possible choices of the background U(1) field's phase:
$\cos\Phi=x^2=\pm 1$. In that  limit eq.\ (\ref{cr1}) becomes
\be
(1-\cos\Phi)=\frac{1}{4t^5}.
\ee
   
In both cases we find that the spectrum 
\be
\cosh\varepsilon(p)=1-2(\cos 2p-1)\frac{1}{t}+
\left(\cos 4p+4\cos 2p-5\right)\frac{1}{t^2}
 \ee
when $t \to \infty$ becomes $\varepsilon(p)=0$
 for the all values of the momentum.
\section{Representation of the partition function $Z$ via two- and
three-particle $R$-matrices}
\indent

In this section we will demonstrate that the partition function
(\ref{Z1}) can be constructed as a trace of the $N$-th power
($N$ is a size of the lattice in a time direction) of some transfer 
matrix $T$
\be
\label{TN}
Z=tr T^N,
\ee
but unlike the case of ordinary integrable models, the transfer
matrix $T$ here is a product of four transfer matrices  presented
in the formulas (\ref{T4}), (\ref{Hi}) and (\ref{4H}).
However, if we  look at the Fig.3 under a  $45^{\circ}$ angle and use
the corresponding time direction, we  see 
only two different rows of products of  constituent
$R$-matrices (see Fig.4), corresponding to following Monodromy matrices
in the so called braid formalism
\bea
\label{RR}
T_0(u)&=&\prod_{j=0}^{L} R^{(2)}_{3j-1,3j}(u) R^{(3)}_{3j,3j+1.3j+2}(u),\\
T_1(u)&=&\prod_{j=0}^{L} R^{(3)}_{3j,3j+1,3j+2}(u) R^{(2)}_{3j+2,3j+3}(u).
\ena
Here the $R^{(2)}_{i,i+1}(u)$'s are the ordinary two-particle scattering 
$R$-matrices
corresponding to the squares of the chain in the Fig.\ 4, and 
the $R^{(3)}_{i-1,i,i+1}(u)$'s  correspond to the hexagons and represent
three-particle scattering $R$-matrices. Therefore the Monodromy matrix of the
model is the product $T(u)= T_1(u) T_0(u)$.

The two-particle $R_{ij}$-matrix is an operator acting on the
 direct product of the two two dimensional spaces
  (spaces of the fermions with 0 spin) defined on the sites i and j,
which according to the technique developed in  \cite{AK}
can be fermionised by considering
the basis states as $|k\rangle$, $k=0,1$, with  $|1\rangle=c^+|0\rangle$:
 \be
 R^{(2)}_{ij}|i_1\rangle\otimes|j_1\rangle=(R_{ij})^{i_2 j_2}_{i_1 
j_1}|i_2\rangle\otimes|j_2\rangle. 
 \ee 
Graphically it is represented in Fig.5b, in the same way as in
\cite{APSS} for
the $R$-matrices of the $XXZ$-model in the braid formalism,

The three-particle $R_{ijk}$-matrix is an endomorphism on the direct 
product of three two-dimensional spaces (see Fig.5a) with a basis defined 
as before:
 \be
 R^{(3)}_{ijk}|i_1\rangle\otimes|j_1\rangle\otimes|k_1\rangle=(R_{ijk})^{i_2 
j_2 k_2}_{i_1 j_1 k_1}
 |i_2\rangle\otimes|j_2\rangle\otimes|k_2\rangle. 
 \ee 

 Following to the \cite{AK} we can represent $R^{(s)}$-matrices $(s=2,3)$
via fermionic creation-annihilation operators by considering the Hubbard 
operators
 \be
 X^{i}_{j}=|j\rangle\langle i|,\qquad i,j=0,1,
 \ee
 and taking into account that the fermionic Fock space is graded with the
parities of the states defined as $p(i)=i$. Then, by definition
\be
 \ba{ll}
 R^{(2)}_{ij}&=R_{ij}|i_1\rangle|j_1\rangle\langle j_1|\langle i_1|
 =(R_{ij})^{i_2 j_2}_{i_1 j_1}|i_2\rangle|j_2\rangle\langle j_1|\langle i_1|\\
 { }&=(-1)^{p(i_1)(p(j_1)+p(j_2))}
(R_{ij})^{i_2 j_2}_{i_1 j_1} X^{i_1}_{i_2} X^{j_1}_{j_2},
 \ea
 \ee
 and
 \be
 \ba{ll}
 R^{(3)}_{ijk}&=R_{ijk}|i_1\rangle|j_1\rangle|k_1\rangle\langle k_1|\langle 
j_1|\langle i_1|\\
 { }&=(R_{ijk})^{i_2 j_2 k_2}_{i_1 j_1 
k_1}|i_2\rangle|j_2\rangle|k_2\rangle\langle k_1|\langle j_1|\langle i_1|\\
 { }&=(-1)^{p(i_1)(p(j_2)+p(j_1))+(p(i_1)+p(j_1))(p(k_1)+p(k_2))}
(R_{ijk})^{i_2 j_2 k_2}_{i_1 j_1 k_1} X^{i_1}_{i_2} X^{j_1}_{j_2} 
X^{k_1}_{k_2}.
 \ea
 \ee
 
The fermionic expression of $R^{(2)}$  for the $XXZ$ model
can be found in \cite{AK}(and references there), while the most general 
form of the three-particle scattering $R^{(3)}_{123}$ operator is
\be
\label{3R}
\ba{ll}
R^{(3)}_{123}=&R_{000}^{000}n_1 n_2 n_3+R_{001}^{001}n_1n_2\bar n_3+
R_{010}^{010}n_1
\bar n_2n_3+
R_{100}^{100}\bar n_1n_2n_3+
R_{011}^{011}n_1\bar n_2\bar n_3\\
&+R_{101}^{101}\bar n_1n_2\bar n_3+
R_{110}^{110}\bar n_1\bar n_2n_3+R_{111}^{111}\bar n_1\bar n_2\bar n_3+
(R_{001}^{010}c^{+}_2c_3
+R_{010}^{001}c^{+}_3c_2)n_1\\
&+(R_{101}^{110}c^{+}_2 c_3+R_{110}^{101}c^{+}_3c_2)\bar n_1+
(R_{001}^{100}c^{+}_1c_3+R_{100}^{001}c^{+}_3 c_1)n_2+(R_{011}^{110}
c^{+}_1 c_3+\\
&R_{110}^{011}c^{+}_3 c_1)\bar n_2+(R_{010}^{100}c^{+}_1  c_2+R_{100}^{010}
c^{+}_2 c_1)n_3+
(R_{011}^{101}c^{+}_1  c_2+R_{101}^{011}c^{+}_2 c_1)\bar n_3,
\ea
\ee
Where $n_i=c^{+}_ic_i$ and $\bar n_i=1-n_i$. The two-particle 
$R^{(2)}_{ij}$-matrix of the $XXZ$ model can be obtained from (\ref{3R})
by putting $\bar n_3=1$ 
everywhere and taking $(R^{(2)})_{i_1j_1}^{i_2j_2}=
(R^{(3)})_{i_1j_1 1}^{i_2j_2 1}$ and
$R_{i_1j_10}^{i_2j_20}=0$ otherwise.

Both kinds of $R$-matrices can be written as an exponent 
\be
\label{A}
:\exp{c^{+}_i A_{ij}c_j}:,
\ee
where the indices $i,j$ run from 1 to 2 for $R^{(2)}_{12}$ operators and to 3 
for the three-particle $R^{(3)}_{123}$ operator. The notion $:\quad :$ 
means the 
normal ordering for the $c^+_i$ , $c_i$ (i=1,3) and the hole
ordering for the  $c^+_2$ , $c_2$ operators. 

The connection between the
matrix elements $A_{i,j}$ and the hopping parameters $t_{i,j}$ in the 
action of the model can be found by the relation
\be
\label{2A}
\langle\bar{\psi}_1|\langle{\psi}_2|\langle\bar{\psi}_3|:\exp{c^{+}_i 
A_{ij}c_j}:
|{\psi}_3\rangle|\bar{\psi}_2\rangle|{\psi}_1\rangle=\exp{(-\bar{\psi}_it_{i,j}{\psi}_j)}.
\ee
and are
\begin{eqnarray}
\label{3A}
 A_{ij}=\left\{\ba{lll}-t_{ij},&i \not = j,& i=1,3\\
 \hspace{3mm}t_{ij}&&i=2
   \ea \right.,    
 A_{ii}=\left\{\ba{lll}\hspace{3mm}1+t_{ii}&i=2&\\
  -(1+t_{ii})&i=1,3&
\ea \right. .
\end{eqnarray}
In (\ref{2A}) the 
coherent states  $|\psi_i\rangle$ are
defined by the formulas (\ref{coherent}).

 The matrix elements $(R^{(3)})_{ijk}^{lmn}$ in the expression (\ref{3R})
are connected with the $A_{ij}$-s in (\ref{A})
by the following equations
\bea
\label{1RA}
\hspace{-17mm}
\begin{array}{ll}
R_{101}^{101}=1,&R_{101}^{110}=A_{23},\hspace{5mm} R_{110}^{101}=A_{32},\\
R_{011}^{011}=(1+A_{11})(1-A_{22})+A_{12}A_{21},&R_{011}^{101}=A_{12},
\hspace{5mm}R_{101}^{011}=A_{21},\\
R_{110}^{110}=(1+A_{33})(1-A_{22})+A_{23}A_{32},&R_{011}^{110}=A_{13}(1-
A_{22})+A_{12}A_{23},\\
R_{111}^{111}=1-A_{22},&R_{110}^{011}=A_{31}(1-A_{22})+A_{21}A_{32},\\
\end{array}
\ena

\be
\label{2RA}
\hspace{-20mm}
\begin{array}{ll}
R_{001}^{010}=A_{23}(1+A_{11})-A_{13}A_{21},&\hspace{11mm}R_{010}^{001}=
A_{32}(1+A_{11})-A_{31}A_{12},\\
R_{001}^{100}=-A_{13}&\hspace{11mm}R_{100}^{001}=-A_{31},\\
R_{010}^{100}=A_{12}(1+A_{33})-A_{32}A_{13},&\hspace{11mm}R_{100}^{010}=
A_{21}
(1+A_{33})-A_{23}A_{31},\\
\end{array}
\ee
\bea
\label{3RA}
\begin{array}{l}
R_{001}^{001}=1+A_{11},\\
R_{010}^{010}=(1-A_{22})(1+A_{11}+A_{33})-\det A+A_{11}A_{33}-A_{13}A_{31}+
A_{12}A_{21}+A_{23}A_{32},\\
R_{100}^{100}=1+A_{33},\\
R_{000}^{000}=(1+A_{11})(1+A_{33})-A_{13}A_{31}.\\
\end{array}
\ena
The two-particle $R^{(2)}_{ij}$-matrix
 can be obtained from these expressions by taking  $A_{i3}=A_{3j}=0$ 
everywhere.

This is the general form of three-particle $R^{(3)}_{ijk}$ matrix, but for the
model under consideration (\ref{Z1}) we should consider free fermionic limit
and take $A_{22}=1$ and $A_{13}=A_{13}=0$.

The condition of integrability of this model with two- and three-particle 
$R$-matrices, namely the condition of
commutativity of Transfer matrices (\ref{RR}) for different values of the 
spectral parameter, can be written as a modified Yang-Baxter equation
in the following form 
 \be
 {\bf R}_{12}(u,v) R_{234}(u) R_{45}(u) R_{12}(v) R_{234}(v) =
R_{234}(v) R_{45}(v) R_{12}(u) R_{234}(u) {\bf R}_{45}(u,v),
 \ee
were ${\bf R}_{ij}(u,v)$ is the intertwiner operator. This equations definitely
has a solution in free fermionic case corresponding to the model defined
above, since we were able to diagonalise the Hamiltonian. It would be
interesting to find a solution for a general case of $R^{(3)}$ matrix
(\ref{1RA}-\ref{3RA}) and $R^{(2)}$-matrix of the $XXZ$ model. 

\section{Acknowledgement}
\indent

The article is devoted to the 70 year Jubelee of Prof. S.G.Matinuyan.

The authors Sh.K. and A.S. acknowledge INTAS grant 00-390 for partial
financial support.

\newpage
\section*{Figure captions}
\indent
1a. The BCC lattice\\
1b. The Dual BCC lattice\\
\\
2.  How the Manhattan Lattice is forming on surfaces on Dual BCC lattice\\
\\
3.  Manhattan lattice appeared on 2D surfaces on Dual BCC lattice\\
\\
4.  Two rows of Monodromy matrices\\
\\
5a. Graphical representation of $R^{(2)}_{ij}$ matrix\\
5b. Graphical representation of $R^{(3)}_{ijk}$ matrix

\newpage
\unitlength=9pt


\begin{picture}(25,25)
\unitlength=6pt
\linethickness{0.5pt}
\newsavebox{\xoranard}
\sbox{\xoranard}{\begin{picture}(16,17)
\multiput(5,0)(12,0){2}{\line(0,1){12}}
\put(5,12){\line(1,0){12}}
\put(17,12){\line(-5,4){5}}
\multiput(12,4)(0,0.5){24}{\circle*{0.15}}
\multiput(0,4)(0.5,0){24}{\circle*{0.15}}
\put(0,4){\line(5,-4){5}}
\multiput(12,4)(0.5,-0.4){10}{\circle*{0.15}}
\multiput(0,4)(12,0){2}{\circle*{1}}
\multiput(5,0)(12,0){2}{\circle*{1}}
\multiput(0,16)(12,0){2}{\circle*{1}}
\multiput(5,12)(12,0){2}{\circle*{1}}
\end{picture}}

\multiput(0,16)(8.5,-8){2}{\usebox{\xoranard}}
\put(0,20){\line(0,1){12}}
\put(0,32){\line(1,0){12}}
\put(0,32){\line(5,-4){5}}
\put(13.5,8){\line(1,0){12}}
\put(5,16){\line(1,0){6}}
\put(8.5,12){\line(0,1){4}}
\put(20.5,24){\line(-1,0){3.5}}
\put(13.5,20){\line(-5,4){2.5}}
\multiput(8.5,24)(0,-0.5){17}{\circle*{0.15}}
\multiput(8.5,24)(0.5,0){17}{\circle*{0.15}}
\multiput(8.5,24)(0.5,-0.4){8}{\circle*{0.15}}
\multiput(11,16)(0.5,0){12}{\circle*{0.15}}

\put(6,0){{\mbox a)}}
\end{picture}

\begin{picture}(18,17)(-28,-20)
\multiput(0,0)(16,0){2}{\line(0,1){13}}
\multiput(1,8)(14,0){2}{\line(0,1){4}}
\multiput(8,4)(3,3){2}{\line(-1,1){3}}
\multiput(8,4)(-3,3){2}{\line(1,1){3}}
\multiput(1,8)(3,7){2}{\line(4,-1){4}}
\multiput(15,8)(-3,7){2}{\line(-4,-1){4}}
\put(1,12){\line(1,1){3}}
\put(15,12){\line(-1,1){3}}
\put(4,15){\line(4,1){4}}
\put(12,15){\line(-4,1){4}}
\put(0,13){\line(1,4){1}}
\put(16,13){\line(-1,4){1}}
\put(1,17){\line(1,0){14}}
\put(1,8){\line(1,-2){1}}
\put(15,8){\line(-1,-2){1}}
\put(2,6){\line(3,-2){3}}
\put(14,6){\line(-3,-2){3}}
\put(8,3){\line(-3,1){3}}
\put(8,3){\line(3,1){3}}
\put(8,3){\line(0,1){1}}
\put(0,0){\line(1,0){16}}
\put(0,13){\line(1,0){16}}
\multiput(0,0)(0.2,0.2){25}{\circle*{0.1}}
\multiput(16,0)(-0.2,0.2){25}{\circle*{0.1}}
\multiput(0,13)(0.3,-0.15){14}{\circle*{0.1}}
\multiput(16,13)(-0.3,-0.15){14}{\circle*{0.1}}
\multiput(1,17)(0.2,-0.3){11}{\circle*{0.1}}
\multiput(15,17)(-0.2,-0.3){11}{\circle*{0.1}}

\put(8,10){\line(0,1){4}}

\multiput(0,0)(16,0){2}{\circle*{1}}
\multiput(0,13)(16,0){2}{\circle*{1}}
\multiput(1,17)(14,0){2}{\circle*{1}}
\multiput(5,5)(6,0){2}{\circle*{0.5}}
\multiput(4,11)(8,0){2}{\circle*{0.5}}

\newsavebox{\ketgitsh}
\sbox{\ketgitsh}{\begin{picture}(2,6)
\multiput(0,0)(0.5,1.5){4}{\line(1,3){0.4}}
\end{picture}}
\put(0,0){\usebox{\ketgitsh}}

\newsavebox{\ketgitshg}
\sbox{\ketgitshg}{\begin{picture}(2,6)
\multiput(2,0)(-0.5,1.5){4}{\line(-1,3){0.4}}
\end{picture}}
\put(14,0){\usebox{\ketgitshg}}

\multiput(1,17)(0.26,-1.56){3}{\line(1,-6){0.2}}
\multiput(15,17)(-0.26,-1.56){3}{\line(-1,-6){0.2}}

\end{picture}

\begin{picture}(80,-20)(-35,-18)
\put(0,0){{\mbox b)}}
\end{picture}

\begin{picture}(50,0)(0,-18)
\put(18,0){{\mbox Fig.1}}
\end{picture}


\begin{picture}(18,17)(-11,-8)
\multiput(1,8)(14,0){2}{\line(0,1){4}}
\multiput(8,4)(3,3){2}{\line(-1,1){3}}
\multiput(8,4)(-3,3){2}{\line(1,1){3}}
\multiput(1,8)(3,7){2}{\line(4,-1){4}}
\multiput(15,8)(-3,7){2}{\line(-4,-1){4}}
\put(1,12){\line(1,1){3}}
\put(15,12){\line(-1,1){3}}
\put(4,15){\line(4,1){4}}
\put(12,15){\line(-4,1){4}}
\put(1,8){\line(1,-2){1}}
\put(15,8){\line(-1,-2){1}}
\put(2,6){\line(3,-2){3}}
\put(14,6){\line(-3,-2){3}}
\put(8,3){\line(-3,1){3}}
\put(8,3){\line(3,1){3}}
\put(8,3){\line(0,1){1}}

\put(8,10){\line(0,1){4}}
\put(6,14,5){\vector(-4,-1){3.2}}
\put(6,14.5){\vector(2,-3){2}}
\put(1,10){\vector(1,2){1.8}}
\put(1,10){\vector(2,-3){1.6}}
\put(6.5,8.5){\vector(1,2){1.5}}
\put(6.5,8.5){\vector(-4,-1){3.7}}
\put(6,15.5){\vector(1,0){3.8}}
\put(6,15.5){\vector(0,-1){0.9}}
\put(10,14.5){\vector(0,1){0.9}}
\put(10,14.5){\vector(-1,0){3.8}}
\put(8,11.5){\vector(2,3){2}}
\put(8,11.5){\vector(1,-2){1.5}}
\put(13.2,13.7){\vector(-4,1){3.2}}
\put(13.2,13.7){\vector(1,-2){1.8}}
\put(13.2,7.6){\vector(2,3){1.8}}
\put(13.2,7.6){\vector(-4,1){3.7}}
\put(9.5,8.5){\vector(-1,0){3}}
\put(9.5,8.5){\vector(0,-1){3}}
\put(6.5,5.5){\vector(1,0){3}}
\put(6.5,5.5){\vector(0,1){3}}
\put(9.5,5.5){\vector(2,1){3.8}}

\end{picture}

\begin{picture}(50,0)(0,-10)
\put(18,0){{\mbox Fig.2}}
\end{picture}

\newpage

\unitlength=5pt

\newsavebox{\bloka}

\sbox{\bloka}{\begin{picture}(6,6)

\put(0,0){\vector(1,0){6}}

\put(6,6){\vector(-1,0){6}}

\end{picture}}

\newsavebox{\blokb}

\sbox{\blokb}{\begin{picture}(6,6)

\put(6,0){\vector(-1,0){6}}

\put(0,6){\vector(1,0){6}}

\end{picture}}

\newsavebox{\blokc}

\sbox{\blokc}{\begin{picture}(6,6)

\put(0,0){\vector(0,1){6}}

\put(6,6){\vector(0,-1){6}}

\end{picture}}

\newsavebox{\blokd}

\sbox{\blokd}{\begin{picture}(6,6)

\put(0,6){\vector(0,-1){6}}

\put(6,0){\vector(0,1){6}}

\end{picture}}

\newsavebox{\bloke}

\sbox{\bloke}{\begin{picture}(6,12)

\put(0,0){\vector(1,2){3}}

\put(3,6){\vector(1,-2){3}}

\put(0,12){\vector(1,-2){3}}

\put(3,6){\vector(1,2){3}}

\end{picture}}

\newsavebox{\blokf}

\sbox{\blokf}{\begin{picture}(6,12)

\put(6,0){\vector(-1,2){3}}

\put(3,6){\vector(-1,-2){3}}

\put(6,12){\vector(-1,-2){3}}

\put(3,6){\vector(-1,2){3}}

\end{picture}}

\begin{picture}(100,50)

 \multiput(0,0)(6,0){10}{\usebox{\blokb}}

\multiput(0,36)(6,0){10}{\usebox{\blokb}}

 \multiput(0,18)(6,0){10}{\usebox{\bloka}}

\multiput(6,0)(12,0){5}{\usebox{\blokc}}

 \multiput(0,18)(12,0){5}{\usebox{\blokc}}

\multiput(6,36)(12,0){5}{\usebox{\blokc}}

 \multiput(0,24)(12,0){5}{\usebox{\bloke}}

\multiput(0,6)(12,0){5}{\usebox{\blokf}}



\put(12.5,0.5){\scriptsize 1}

\put(18.5,0.5){\scriptsize 2}

\put(18.5,6.4){\scriptsize 3}

\put(12.7,6.4){\scriptsize 4}

\put(15.7,11.8){\scriptsize 5}

\put(18.5,18.5){\scriptsize 6}

\put(12.5,18.5){\scriptsize 7}

\put(12.8,24.4){\scriptsize 8}

\put(18.5,24.4){\scriptsize 9}

\put(15.5,30){\scriptsize 0}

\put(68,0){\vector(1,0){5}}

\put(68,0){\vector(0,1){5}}

\put(72,0.5){\small x}

\put(68.5,4.5){\small y}

\put(61,0.5){\small t}

\put(61,6.5){\small $t+1$}

\put(61,18.5){\small $t+2$}

\put(61,24.5){\small $t+3$}

\put(26.5,20){\small $B_1$}
\put(26.5,2.5){\small $B_2$}
\put(20,11){\small $H_1$}
\put(20,29){\small $H_2$}
\put(20.5,2.5){\small $A_1$}
\put(20.5,20){\small $A_1$}

\end{picture}

\begin{picture}(50,1)(0,1)
\put(28,0){{\mbox Fig.3}}
\end{picture} 

\newpage
\unitlength=5pt

\newsavebox{\blokh}

\sbox{\blokh}{\begin{picture}(12,12)
\put(3,0){\vector(1,0){6}}
\put(3,0){\vector(-1,2){3}}
\put(3,12){\vector(-1,-2){3}}
\put(3,12){\vector(1,0){6}}
\put(12,6){\vector(-1,-2){3}}
\put(12,6){\vector(-1,2){3}}
\end{picture}}

\newsavebox{\blokho}

\sbox{\blokho}{\begin{picture}(12,12)
\put(9,0){\vector(-1,0){6}}
\put(0,6){\vector(1,-2){3}}
\put(0,6){\vector(1,2){3}}
\put(9,12){\vector(-1,0){6}}
\put(9,0){\vector(1,2){3}}
\put(9,12){\vector(1,-2){3}}
\end{picture}}

\newsavebox{\blokq}

\sbox{\blokq}{\begin{picture}(6,12)
\put(3,0){\vector(-1,2){3}}
\put(3,0){\vector(1,2){3}}
\put(3,12){\vector(-1,-2){3}}
\put(3,12){\vector(1,-2){3}}
\end{picture}}

\newsavebox{\blokqt}

\sbox{\blokqt}{\begin{picture}(6,12)
\put(0,6){\vector(1,-2){3}}
\put(0,6){\vector(1,2){3}}
\put(6,6){\vector(-1,-2){3}}
\put(6,6){\vector(-1,2){3}}
\end{picture}}

\begin{picture}(100,24)
\multiput(0,0)(36,0){2}{\usebox{\blokh}}
\multiput(6,12)(36,0){2}{\usebox{\blokh}}
\multiput(18,0)(36,0){2}{\usebox{\blokho}}
\multiput(24,12)(36,0){2}{\usebox{\blokho}}
\multiput(18,12)(36,0){2}{\usebox{\blokq}}
\multiput(30,0)(36,0){2}{\usebox{\blokqt}}
\multiput(0,12)(36,0){2}{\usebox{\blokqt}}
\multiput(12,0)(36,0){2}{\usebox{\blokq}}
\put(-4,17){$T_0$}
\put(-4,5){$T_1$}
\end{picture}

\begin{picture}(100,2)
\put(3,0){1}
\put(9,0){2}
\put(15,0){3}
\put(21,0){4}
\put(27,0){5}
\put(32,0){6}
\end{picture}

\begin{picture}(50,6)
\put(34,0){Fig.4}
\end{picture}

\begin{picture}(25,30)(-5,0)
\put(3,3){\vector(1,0){6}}
\put(3,3){\vector(-1,2){3}}
\put(3,15){\vector(-1,-2){3}}
\put(3,15){\vector(1,0){6}}
\put(12,9){\vector(-1,2){3}}
\put(12,9){\vector(-1,-2){3}}
\put(5,9){\small{\mbox H}}
\put(14,9){$=R^3_{123}$}

\put(1.6,2.5){\scriptsize 2}
\put(9.9,2.5){\scriptsize 3}
\put(2,15.2){\scriptsize 1}
\put(-1.2,8.5){\scriptsize 1}
\put(12.6,8.5){\scriptsize 3}
\put(9.2,15.2){\scriptsize 2}
\put(15,1){\small a)}
\end{picture}

\begin{picture}(25,30)(-25,-30)
\put(25,6){\vector(1,0){7}}
\put(25,6){\vector(0,1){7}}
\put(32,13){\vector(-1,0){7}}
\put(32,13){\vector(0,-1){7}}

\put(24.8,4.6){\scriptsize 1}
\put(24.8,13.5){\scriptsize 1}
\put(32,4.6){\scriptsize 2}
\put(32,13.5){\scriptsize 2}
\put(28,8.5){\small{\mbox B}}
\put(34,9){$=R^2_{12}$}
\put(35,1){\small b)}
\end{picture}

\begin{picture}(50,2)(0,-25)
\put(34,0){{\mbox Fig.5}}
\end{picture}

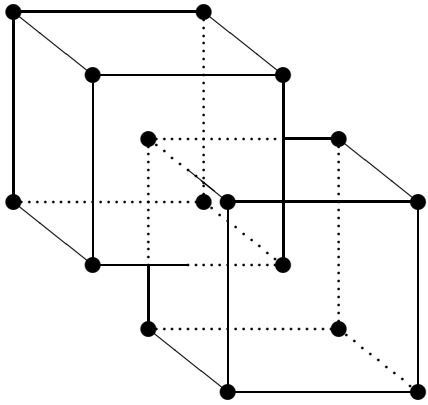
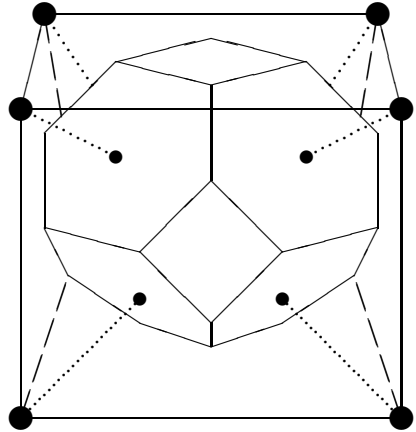
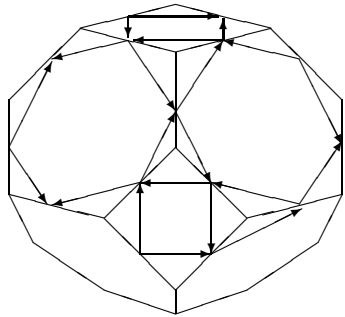
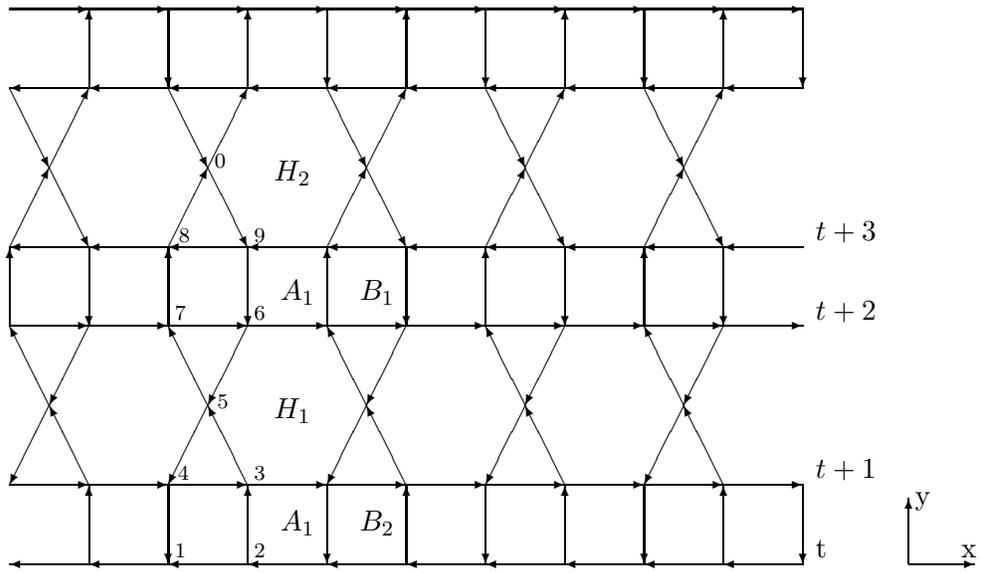
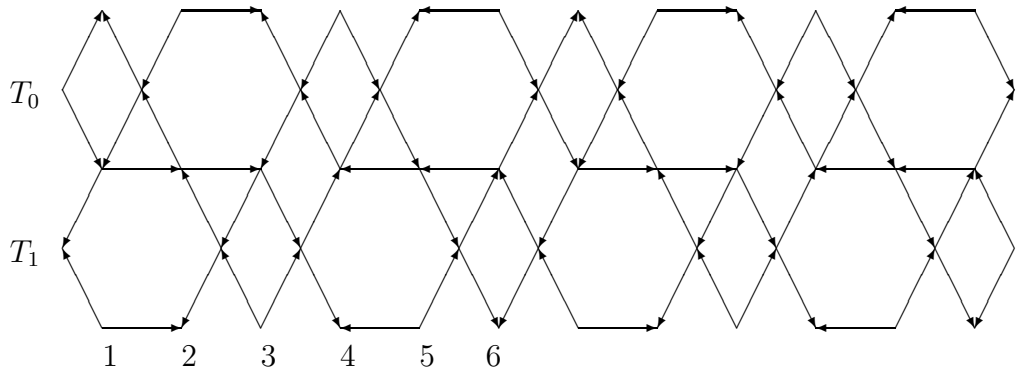
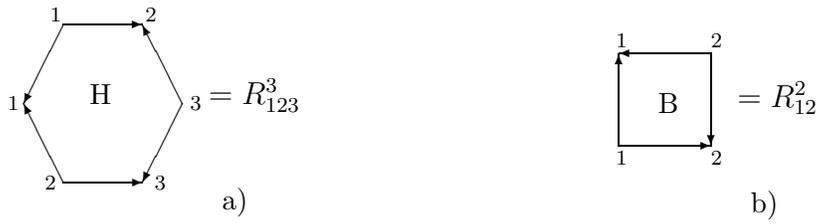
\end{document}